\documentstyle[preprint,aps,epsf,eqsecnum]{revtex}
\begin{document}
\preprint{\vbox{\hbox{BARI-TH/97-265 \hfill} 
                \hbox{April 1997 \hfill}}}
\title{Investigations on CPT Invariance at $B$-Factories}
\author{P. Colangelo $^1$ and G. Corcella $^2$}
\address{
$^1$ I.N.F.N., Sezione di Bari, Italy\\
$^2$ Dipartimento di Fisica dell'Universit\'a di Milano, and\\ 
I.N.F.N., Sezione di Milano, Italy\\
}
\maketitle
\begin{abstract}
We study the sensitivity of an asymmetric $B$-factory to indirect $CPT$ 
violating effects. We find that both dilepton signals and nonleptonic 
asymmetries can be used to bound the space of parameters describing 
possible $CPT$ violation in the mixing matrix of the neutral $B$ system.
\end{abstract}

\newpage
\section{Introduction}
Invariance under $CPT$ transforms is an exact property of a relativistic, local 
and renormalizable field theory, satisfying causality and invariance under 
proper Lorentz transformations, incorporating usual spin-statistic relations 
and provided with asymptotic states \cite{schwinger}.
Therefore, tests of $CPT$ invariance probe fundamental aspects of the current 
field-theoretical description of microscopic phenomena, and provide stringent 
constraints on quantum gravity, supergravity and superstring-inspired scenarios
requiring a violation of $CPT$ \cite{hawking}.

$CPT$ symmetry, which implies the equality of mass, lifetime, branching ratios 
of particles and antiparticles, was tested at various levels of accuracy 
since the early days of the high-energy physics era \cite{pdg}.
The main activity was focused on the theoretical and experimental 
analysis of the neutral kaon system, where it was observed that the phases 
$\phi_{+-}$, $\phi_{00}$ and $\phi_{SW}$ are equal within an accuracy of 
$1^\circ$ \cite{cplear}.
At $\Phi$-factories, where $8.9\times 10^9$ $K_L-K_S$ pairs per year 
will be produced, tests of even higher accuracy are expected to 
be performed in the near future\cite{maiani}.

Tests of $CPT$ invariance have been proposed for the neutral $D$ 
meson system \cite{colladay}. 
Moreover, the symmetric and asymmetric $B$-factories currently
under construction \cite{babar,belle,cleo3} represent experimental 
facilities where 
$CPT$ symmetry can be investigated in the neutral $B$ system.
In this case, information is used from $\Upsilon (4S)$ decays to coherent
$C$-odd $B^0$-$\bar B^0$ states ($C$-even states, e.g. from  
$\Upsilon(4S)\to B^0\bar B^0\gamma$, are expected at the level of $1/10^9$  
events \cite{corcella})
so that clear $CPT$ tests, in addition to $CP$ measurements, can be 
carried out.

The proposal of using the neutral $B$ system for testing $CPT$ was already put 
forward in refs. \cite{sanda,kostelecky}. In particular, in ref.
\cite{kostelecky} the features of symmetric and asymmetric $B$-factories were
considered in order to put bounds to possible violations 
of $CPT$, mainly using the channel where one neutral $B$ meson decays 
{\it via} a 
semileptonic transition, and the other one decays to
$J/\psi K_S$; Monte Carlo simulations were used 
to study the sensitivity to this process in various
experimental environments.
In this paper we want to reconsider the problem.
In particular, we investigate the feasibility of bounding $CPT$ invariance 
in the neutral $B$ system at an asymmetric $B$-factory using not only 
nonleptonic decay channels, but also the decay modes where both the neutral $B$ 
mesons decay via  a semileptonic transition, 
since a high statistics is expected in this case, 
with limited reconstruction and background rejection difficulties. 
We introduce a $CPT$ violating term in the mixing matrix of neutral $B$ mesons, 
according to the notation in ref.\cite{sanda}, and then we consider the case 
where lepton pairs are identified in semileptonic $B^0$ and $\bar B^0$ decays.

We also consider the decay channel $B\to J/\psi K_S$ and report on the 
sensitivity to different decay modes, such as $B\to D^+D^-$ and 
$B\to {D^*}^+{D^*}^-$.

Our conclusion is that it is possible to sensibly
constrain the parameter space of 
$CPT$ violating effects in the neutral $B$ mixing matrix. The future 
$B$-factories, therefore, represent powerful facilities for probing a
fundamental aspect of the field theoretical description of the elementary 
interactions.

\section{Parameterizing $CPT$ violation in the $B^0 - \bar B^0$ mixing matrix}

In the Wigner-Weisskopf approach, the mixing $B^0$-$\bar B^0$ is governed by 
the Hamiltonian
\begin{equation}
H= M-i{\Gamma\over 2}=\left( \matrix{ M_{11}-i{\Gamma_{11}\over 2} & 
M_{12}-i{\Gamma_{12}\over 2}\cr M_{12}^*-i{\Gamma_{12}^*\over 2} &
M_{22}-i{\Gamma_{22}\over 2}\cr} \right) \, \label{ham}
\end{equation}
\noindent
acting on the vector 
$\left( \matrix{ B^0 \cr \bar B^0\cr} \right)$ \ .
The condition 
$H_{11}\not=H_{22}$, or $<B^0|H|B^0>\  \not= \  <\bar B^0|H|\bar B^0> \ ,$
implies an indirect violation of $CPT$ invariance in the neutral $B$ system.
Following \cite{sanda} and \cite{wu}, such a violation can be easily 
parameterized expressing the matrix (\ref{ham}) as follows:
\begin{equation}
H= -iD+\sigma_1E_1+\sigma_2E_2+\sigma_3E_3 \label{param}
\end{equation}
with $\sigma_i$ the Pauli matrices and $D, E_1, E_2, E_3$ four complex numbers.
The correspondence between the parameters in eqs.(\ref{ham}-\ref{param}) 
is simply given by:
\begin{eqnarray}
E_1&=&ReM_{12}-{i\over 2}Re\Gamma_{12} \nonumber \\
E_2&=&-ImM_{12}+{i\over 2} Im\Gamma_{12} \nonumber \\
E_3&=&{1\over 2} (M_{11}-M_{22})-{i\over 4} (\Gamma_{11}-\Gamma_{22}) \ .
\end{eqnarray}
$CP$ and $CPT$ violating effects can be
parameterized \cite{sanda,wu} by introducing the new complex variables $E$, 
$\theta$, $\phi$, defined as:
\begin{equation}
E={(E_1^2+E_2^2+E_3^2)}^{1/2} \ , \ E_1=E\sin\theta\cos\phi\ , \
E_2=E\sin\theta\sin\phi \ , \ E_3=E\cos\theta \;\;\;.
\end{equation}
As a matter of fact, $CPT$ symmetry implies 
$M_{11}=M_{22}$\  and \ $\Gamma_{11}=\Gamma_{22}$, i.e. $\cos\theta=0$.
Moreover, using the phase convention
$CP|B^0>=|\bar B^0>$, $CP$ invariance requires the equality
$<B^0|H|\bar B^0>=<\bar B^0|H| B^0>$, i.e.:
\begin{equation}
M_{12}-i {\Gamma_{12} \over 2}=M_{12}^*-i {\Gamma_{12}^* \over 2} \ .
\end{equation}
This means $Im\; \Gamma_{12}=Im \; M_{12}=0$, i.e. 
$\phi=0$ and $\cos\theta=0$.

The coefficients relating the mass eigenstates of the Hamiltonian (\ref{ham}) 
to the 
flavour eigenstates $|\buildrel{(-)}\over {B^0}>$:
\begin{equation}
|B_1>={1\over {(|p_1|^2+|q_1|^2)^{1/2}}}\left( p_1|B^0>+q_1|\bar B^0>\right)
\end{equation}
\begin{equation}
|B_2>={1\over {(|p_2|^2+|q_2|^2)^{1/2}}}\left( p_2|B^0>-q_2|\bar B^0>\right)
\end{equation}
are simply related to the parameters $\theta$ and $\phi$ by the equations
\begin{equation}
{{q_1}\over {p_1}}=e^{i\phi} \tan{\theta\over 2}\;\;\;\; \;  ,\;\;\;\;\;  
{{q_2}\over {p_2}}=e^{i\phi} \cot{\theta\over 2}\ .
\end{equation}
Then, $CPT$ invariance implies $\theta={\pi \over 2}$ and 
\begin{equation}
{{q_1}\over {p_1}}={{q_2}\over {p_2}}=e^{i\phi}\ ;
\end{equation} 
a violation of $CPT$ is governed by the parameter 
\begin{equation}
s=\cot\theta={1\over 2} {\left({{q_2}\over {p_2}}-{{q_1}\over {p_1}}\right)} 
e^{-i\phi}\ .
\end{equation}
Although, in principle, $\phi$ and $\theta$ are complex numbers, we assume in 
the following
\begin{equation}
Im \; \phi = 0 \; ,\; Im \;\theta = 0 \;\;\; ,  
\end{equation}
and therefore a real variable $s$, to reduce the phenomenological analysis to 
the 
effects of the variation of a single parameter. We shall comment below 
on the possible effects related to the imaginary part of $s$. 

If at $t=0$ only a $B^0$ or $\bar B^0$ state is present, 
the time evolution of the mass eigenstates $|B_1>$ and $|B_2>$:
\begin{equation}
|B_n(t)>={e^{-i(m_n-i{{\Gamma_n}\over 2})t}} |B_n(0)>\; \; \;\;\;\;
  (n=1,2) 
\end{equation}
implies that
\begin{equation}
|B^0(t)>=e^{-(im+{\Gamma\over 2})t}\bigg [g_+(t)|B^0>
+\bar g_+(t)|\bar B^0>\bigg ] 
\label{b0t}
\end{equation}
\begin{equation}
|\bar B^0(t)>=e^{-(im+{\Gamma\over 2})t}[\bar g_-(t)|B^0>
+ g_-(t)|\bar B^0>] \ ,
\end{equation}
where \cite{sanda}                         
\begin{equation}
g_\pm (t)=\cos^2{\theta \over 2}\  e^{\pm(i\Delta m -{{\Delta\Gamma}\over 2})
{t\over 2}}+\sin^2{\theta \over 2}\  e^{\mp(i\Delta m -{{\Delta\Gamma}\over 2})
{t\over 2}}
\end{equation}
\begin{equation}
\bar g_\pm (t)=\sin{\theta\over 2}\cos{\theta\over 2} 
\ [e^{(i\Delta m -{{\Delta\Gamma}\over 2}){t\over 2}}
- e^{-(i\Delta m -{{\Delta\Gamma}\over 2}){t\over 2}}]\ e^{\pm i\phi}\ ;
\label{gpiu}
\end{equation}
\noindent
$m_1(\Gamma_1)$ and $m_2(\Gamma_2)$ are the mass (width) of $|B_1>$ and $|B_2>$,
respectively, and
$m={{m_1+m_2}\over 2}$, $\Delta m=m_2-m_1$,  
$\Gamma={{\Gamma_1+\Gamma_2}\over 2}$, $\Delta\Gamma=\Gamma_1-\Gamma_2$.

\noindent
Since ${{\Delta\Gamma}\over \Gamma} {\buildrel < \over \sim}\ 
10^{-2}$ \cite{bigi}, in the following we neglect the actual value of 
$\Delta\Gamma$.

Eqs.(\ref{b0t}-\ref{gpiu}) show that the time evolution of neutral $B$ states 
is 
governed also by the $CPT$ violating parameter $s$, and then it is possible 
to define observables sensitive to $s$. 
A bound on $s$ can be inferred from the mass difference $B^0$-
$\bar B^0$, using the relation
\begin{equation}
m_{B^0}-m_{\bar B^0}={s\over {\sqrt{1+s^2}}} (m_1-m_2)\ ,
\end{equation}
with $m_1-m_2=3\times 10^{-13}$ $GeV$ \cite{pdg} the measured 
mass difference between $B_1$ and $B_2$.
A value $s\sim {\cal O}(10^{-1})$ implies a relative mass difference 
${{|m_{B^0}-m_{\bar B^0}|}\over {{m_B}_{average}}}\sim {\cal O}(10^{-15})$;
in the neutral kaon system the bound 
${{|m_{K^0}-m_{\bar K^0}|}\over {{m_K}_{average}}}\leq 9\times 10^{-19}$ 
\cite{pdg} corresponds to $s\leq 2\times 10^{-4}$. 

\section{Dileptons}
Let us now consider the signals of a possible 
$CPT$ violation in various final states, starting from the processes where
both the neutral $B$ mesons decay {\it via} a semileptonic transition.

The $B^0\bar B^0$ wave function from $\Upsilon(4S)$ decays reads  
\begin{equation}
|\psi(t)>={1\over {\sqrt 2}} \left[|B^0(\vec k, t)>|\bar B^0(-\vec k, t)>+
C|B^0(-\vec k, t)>|\bar B^0(\vec k, t)>\right]
\end{equation}
where $\pm \vec k$ are the meson momenta in the 
$\Upsilon(4S)$ rest frame, and 
$C=- 1$ is the charge conjugation of the $B^0\bar B^0$ pair.

Let us determine the probability of opposite sign dilepton production 
($l^+X^-$), ($l^-X^+$), $X^{\pm}$ being a generic hadronic state. 
If $t_1$ and $t_2$ are the decay times of the 
$B$ mesons with momentum $\vec k$ and $-\vec k$, respectively, and 
if $P^{\pm\ \mp}(t_1,t_2)$ is the probability for a positive (negative) lepton 
at the time $t_1(t_2)$, one has: 
\begin{eqnarray}
&&P^{-+}(t_1,t_2)\propto|<l^-X^+,t_1;l^+X^-,t_2|\psi>|^2\\ \nonumber
&&\propto{1\over 2}{\bigg |<l^-X^+,t_1|H_W|
B^0(t_1)><l^+X^-,t_2|H_W|\bar B^0(t_2)>}\\ \nonumber
&&-<l^-X^+,t_1|H_W|\bar B^0(t_1)><l^+X^-,t_2|H_W| B^0(t_2)>\bigg |^2\ .
\end{eqnarray}  
Since, within the Standard Model, only the transitions    
$\bar B^0 \to l^- X^+$, $B^0\to l^+ X^-$ occur at the leading order 
in the Fermi constant, the following relations can be derived:
\begin{eqnarray}
<l^+X^-,t|H_W|\bar B^0(t)>&=&e^{-(im+{\Gamma\over 2}){t\over 2}}\ \bar g_-(t)
<l^+X^-|H_W|B^0> \ ,\\
<l^-X^+,t|H_W|B^0(t)>&=&e^{-(im+{\Gamma\over 2}){t\over 2}}\ \bar g_+(t)
<l^-X^+|H_W|\bar B^0> \ ,\\
<l^-X^+,t|H_W|\bar B^0(t)>&=&e^{-(im+{\Gamma\over 2}){t\over 2}}\ g_-(t)
<l^-X^+|H_W|\bar B^0> \ ,\\
<l^+X^-,t|H_W|B^0(t)>&=&e^{-(im+{\Gamma\over 2}){t\over 2}}\ g_+(t)
<l^+X^-|H_W|B^0>\ .  
\end{eqnarray}
\noindent
Neglecting violation of $CPT$ in the decay amplitude, we can put:
\begin{equation}
 <l^+X^-|H_W|B^0>=<l^-X^+|H_W|\bar B^0>=A_l\ . \end{equation}
\noindent
Then, keeping only terms of ${\cal O}(s^2)$, the probability of production of
opposite sign dileptons is given by:
\begin{eqnarray}
P^{-+}(t_1,t_2)&\propto & {1\over 2} e^{-\Gamma (t_1+t_2)}(|A_l|^2)^2
|\bar g_+(t_1)\bar g_-(t_2)
-g_-(t_1)g_+(t_2)|^2=\nonumber \\
&=&{1\over 2}(|A_l|^2)^2 {{e^{-\Gamma (t_1+t_2)}}\over {1+s^2}}
[1+2s^2+\cos \Delta m(t_1-t_2)]\ . \label{pmp}
\end{eqnarray}
The production probability of equal sign dileptons can be derived in an 
analogous way; for $C=-1$ one has  
\begin{equation} 
P^{\pm \pm}(t_1, t_2)\propto (|A_l|^2)^2{{e^{-\Gamma (t_1+t_2)}}\over 
{4(1+s^2)}}[1-\cos \Delta m(t_1-t_2)]\ . \label{ppp}
\end{equation}
For $s=0$, eqs. (\ref{pmp}-\ref{ppp}) reduce to well known expressions for the 
production probability of lepton pairs in $B^0$-$\bar B^0$ decays.
Time integration of (\ref{pmp}) gives the integrated probability
\begin{equation}
P^{-+}\propto {{(|A_l|^2)^2}\over {\Gamma ^2}}\  {1\over {2(1+s^2)}}
\left[1+2s^2+{1\over {1+x_d^2}}\right]\ ,\end{equation}
where $x_d=\displaystyle{{{\Delta m}\over \Gamma}}$\ .
The quantity ${{(|A_l|^2)^2}\over {\Gamma^2 }}$ is proportional to the 
product of the semileptonic $B^0$, $\bar B^0$ branching ratio;
the fractions $n^{\pm \pm}$ of equal and opposite sign dileptons 
produced for each decay can be easily derived:
\begin{equation}
 n^{-+}={1\over {2(1+s^2)}}\left[1+2s^2+{1\over {1+x_d^2}}\right]\ ,
\end{equation}   
and
\begin{equation}
 n^{++}=n^{--}={1\over {4(1+s^2)}} {{x_d^2}\over {1+x_d^2}}\ .
\end{equation}
The dependence on the parameter $s$ signals $CPT$ violation.

A measurement of the ratio
\begin{equation}
R={{n^{++}+n^{--}}\over {n^{+-}}}=
{{x_d^2}\over {1+(1+2s^2)(1+x_d^2)}} \label{R}
\end{equation} 
provides a bound on $s$. Moreover, a measurement of time-dependent 
production, possible at an asymmetric $B$-factory, provides us 
with the fraction of opposite and equal sign dileptons:
\begin{equation}
 n^{-+}(\Delta t)={{e^{-\Gamma |\Delta t|}}\over {2(1+s^2)}} 
[1+2s^2+\cos\Delta m\Delta t] \end{equation}
\begin{equation}
 n^{++}(\Delta t)=n^{--}(\Delta t)=
{{e^{-\Gamma |\Delta t|}}\over 
{4(1+s^2)}}[1-\cos\Delta m\Delta t] \ .
\end{equation} 
Then, the ratio $R(\Delta t)$, analogous to (\ref{R}), can be written as:
\begin{equation}
 R(\Delta t)={{n^{++}(\Delta t)+n^{--}(\Delta t)}\over 
{n^{-+}(\Delta t)}}
={{1-\cos\Delta m\Delta t}\over {1+2s^2+\cos\Delta m\Delta t}}\ .
\label{Rdt}
\end{equation}
\noindent
It is worth observing that other asymmetries, 
obtained reconstructing  the temporal sequences of production of
equal and opposite-sign dileptons:
\begin{equation}
 A={{n^{++}-n^{--}}\over {n^{++}+n^{--}}}\quad , \quad 
\bar A={{n^{-+}-n^{+-}}\over {n^{-+}+n^{+-}}}\ ,\ \end{equation}
vanish (both for positive and negative charge conjugation of the initial $B^0 
\; \bar B^0$ states). However, this is only 
the case for a real $s$. In general, one finds \cite{sanda}:
\begin{equation}
 \bar A(\Delta t)={{n^{-+}(\Delta t)-n^{+-}(\Delta t)}
\over {n^{-+}(\Delta t)+n^{+-}(\Delta t)}}=
-{{2 (Im \; s)\  \sin\Delta m|\Delta t|}\over {1+\cos\Delta m\Delta t}}\ ,
\label{Adt}
\end{equation}
and therefore from the asymmetry (\ref{Adt})
a bound on $Im\;s$ can be derived.

It is now interesting to determine the upper bound on $s$ that is possible 
to obtain in the BaBar experiment at the SLAC $B$-factory PEP II making use of 
dilepton events.
At PEP II $N(B^0\bar B^0)\simeq 1.8\times 10^7$ $B^0\bar B^0$ pairs are 
expected per year, for a machine running at the design luminosity \cite{babar}. 
Using the value of the semileptonic branching fraction  
$B(B^0\to l^+X^-)\simeq 0.105$, and the estimated lepton tagging efficiency 
$\epsilon_{tag}=0.65$ \cite{babar}, the number of expected dilepton events 
per year is:
\begin{equation}
N^{++}=N^{--}=N(B^0\bar B^0)[B(B^0\to l^+X^-)]^2\epsilon_{tag}^2 n^{++}\simeq 2.
8\times 10^4\ ,
\end{equation} 
\begin{equation}
N^{-+}=N(B^0\bar B^0)[B(B^0\to l^+X^-)]^2\epsilon_{tag}^2 n^{-+}\simeq 
2.8\times 10^5 .
\end{equation} 
The error on the variable $R$ can be derived assuming a binomial 
distribution for $N^{\pm\pm}$ and $N^{-+}$.
If $p$ and $1-p$ are the probabilities of two leptons produced with 
opposite and equal charges: 
\begin{equation} 
p=n^{-+}={1\over 2 (1+ s^2)}\left(1+2 s^2+{1\over {1+x_d^2}}\right)\ ,
\end{equation}
\begin{equation} 
1-p=n^{++}+n^{--}={1\over 2 (1 + s^2)}{x_d^2\over {1+x_d^2}}\ 
\end{equation}
(for $s^2<<1$), one has that the probability distribution 
\begin{equation} 
P(N^{-+})= \left(\matrix{\ N\  \cr N^{-+}\cr}\right) 
p^{N^{-+}}(1-p)^{N-N^{-+}}\ ,
\end{equation}
with $N=N^{++}+N^{--}+N^{-+}$, can be expressed in terms of $R$ in the form

\begin{equation} 
P\left({N\over {1+R}}\right)= \left(\matrix{\ N\  \cr {N\over {1+R}}\cr}
\right)p^{N\over {1+R}}\ (1-p)^{N-{N\over {1+R}}}\ ,
\end{equation} 
yielding
\begin{equation} 
\sigma^2\left({N\over {1+R}}\right)=Np(1-p)
\end{equation}
and finally the error $\sigma(R)$:
\begin{equation} 
\sigma (R)=(1+R)^2\sqrt{{2\ N^{-+}\ N^{++}}\over {N^3}}\ .
\end{equation} 
In Fig.1 the ratio $R$ in eq.(3.13) is plotted as a function of $x_d$ and 
$s$ in the ranges $0.63<x_d<0.83$ and $0<s<0.3$.
In Fig.2 we plot the time dependent ratio $R(\Delta t)$ in eq.(3.14).
The corresponding error bars have been calculated using (3.26). For 
time-integrated measurements, the error $\sigma(R)$ is given by 
$\sigma(R)=9\times 10^{-4}$ for $x_d=0.73$, while, considering the whole range 
of $x_d$, $\sigma (R)$ runs from $8\times 10^{-4}$ to $1\times 10^{-3}$.

These errors give an insight on the possible accuracy that can be obtained at 
a $B$-factory such as PEP II, due both to the machine luminosity and to the 
efficient lepton identification. The small value of $\sigma(R)$ shows that 
the region of possible values of $x_d$ and $s$ can be tightly constrained. 
The 
measurements of time integrated and time-dependent dilepton production 
fractions allow to establish an  upper bound to the parameter $s$.
As a matter of fact, from Fig.2 we get that, for $\Delta m\Delta t={\pi\over 
4}$, the bounds $s<(5,7,8) \times 10^{-2}$ are 
obtained, respectively, within one, two and three standard deviations on 
$R(\Delta t)$; for $\Delta m\Delta t={\pi\over 2}$ and ${{3\pi}\over 4}$ the 
bounds are $s<(4,6,7)\times 10^{-2}$, and 
$s<(4,5,6)\times 10^{-2}$, respectively.

\section{Lepton-hadron channels}

\subsection{Decay mode $B\to J/\psi K_S$}

Now we wish to investigate the possibility of bounding $CPT$ violating effects
by means of nonleptonic decays of neutral $B$ mesons
 \cite{kostelecky} , using our parameterization 
of the $CPT$ breaking term.
We consider events where one $B$ decays via a semileptonic transition to
$l^\pm X^\mp$, while the other $B$ decays into a nonleptonic state.
On the experimental side, the semileptonic decay can be used to tag 
the flavour of the other $B$ meson decaying into the hadronic final state $|f>$.

Integrated and time-dependent $CP$ asymmetries of the type
\begin{equation}
A={{N(l^-X^+,f)-N(l^+X^-,\bar f)}\over {N(l^-X^+,f)+N(l^+X^-,\bar f)}}
\label{a1}
\end{equation}
\begin{equation} 
A(\Delta t)={{N(l^-X^+,f;\Delta t)-N(l^+X^-,\bar f;\Delta t)}\over 
{N(l^-X^+,f;\Delta t)+N(l^+X^-,\bar f;\Delta t)}}
\label{a1t}
\end{equation}
can be used to test $CPT$ invariance. 
Let us consider, for example,  the hadronic decay  
$\buildrel {(-)}\over {B^0}\to J/\psi K_S$.
A tree and a penguin diagram contribute to such a process, so that we can 
express the transition amplitudes in the form:
\begin{equation}
<J/\psi K_S|H_W|B^0>=A_T e^{i\delta_T} e^{i\phi_T} + A_P e^{i\delta_P} e^{i\phi_P}
\end{equation}
\begin{equation}
<J/\psi K_S|H_W|\bar B^0>=A_T e^{i\delta_T} e^{-i\phi_T} + 
A_P e^{i\delta_P} e^{-i\phi_P},
\end{equation}
where $A_{T,P}$ and $\phi_{T,P}$ are the weak amplitudes 
and phases respectively, while $\delta_{T,P}$ are the strong phases. 
At the leading order in $s$ the $CP$ asymmetries (\ref{a1}) and (\ref{a1t}) 
read respectively
\begin{equation}
A={{x_d^2}\over {1+x_d^2}}s \cos 2\beta\ ,
\label{abk}
\end{equation}
\begin{equation}
A(\Delta t)=-\sin 2\beta\sin \Delta m\Delta t +s\cos 2\beta (1-\cos \Delta m
\Delta t)\ ,
\label{abkt}
\end{equation}
in terms of the $\beta$ angle of the unitarity triangle. 

The sensitivity of PEP II to a $CPT$ violation effect
in this channel can be obtained 
noticing that the error on $A$ is given by
\begin{equation}
\sigma (A)=d \sqrt{{(1+A)(1-A)}\over {N_{eff}}}\ .
\label{sa}
\end{equation}
In eq.(\ref{sa}) the effective number of events $N_{eff}$ is given by: 
$N_{eff}=N(f)K(f,\bar f)E_{tag}$,
with $N(f)$ the number of events with a reconstructed hadronic state, 
$K(f,\bar f)=n(f,l^+X^-)+n(\bar f,l^-X^+)$, $n(\buildrel{(-)}\over f, 
l^\pm X^\mp)$ being the fractions of lepton-hadron events produced in the 
decay of the $B^0\bar B^0$ pair, $E_{tag}$ the total efficiency of tagging, 
$d$ the dilution factor for the production of $|f>$ states from events 
different from $\buildrel {(-)}\over {B^0} \to f$ decays.
For the final state $f= J/\psi K_S$ the values $E_{tag}=0.34$, $N(f)=886$ and 
$d=1$ are expected \cite{babar}.
In Fig.3 and Fig.4 we plot integrated asymmetries for $\beta = 18^\circ$,
$\beta = 22^\circ$ and $\beta = 26^\circ$,
and time-dependent asymmetries 
for $\beta = 22^\circ$, with the corresponding error bars. 
We notice that the maximum value of the time-dependent asymmetry grows 
with $s$, and therefore 
higher order terms in the $s$-expansion 
are needed to fulfil the bound $|A|\le1$.

For integrated measurements at PEP II, the expected error on $A$ is:  
$\sigma=5.8\times 10^{-2}$, quite  independent of $\beta$. As a result, it is 
possible to 
bound $s\leq 0.21$ for $\beta = 18^\circ$, $s\leq 0.24$ for 
$\beta = 22^\circ$ and $s\leq 0.28$ for $\beta=26^\circ$ within one standard 
deviation.

From time-dependent asymmetries, in the PEP II experimental
environment the bounds
$s\leq 0.13$ ($0.26$) (within one and two standard deviations)
can be established in correspondence to
$\Delta m\Delta t= {\pi\over 2}$, and $s\leq 0.17$ for $\Delta m\Delta 
t={{3\pi}\over 4}$.

\subsection{Other decay modes}
Other hadronic decay modes, such as $B\to D^+D^-$ and $B\to {D^*}^+ {D^*}^-$,
can be used to bound $CPT$ invariance. To calculate the asymmetries
for such final states, new information 
has to be considered concerning the transition amplitude and the final state 
interaction.

The transition amplitudes $B \to D^{(*)} D^{(*)}$ 
are governed by the effective hamiltonian \cite{buras}
\begin{equation}
H_{eff}(\Delta B=-1)={G\over {\sqrt 2}}[V_{ub}V^*_{uq}(c_1O^u_1+c_2O^u_2)+
V_{cb}V^*_{cq}(c_1O^c_1+c_2O^c_2)-V_{tb}V^*_{tq}\sum_{i=3}^6{c_iO_i}]\ ,
\end{equation}
where $V_{jb}$ and $V^*_{uq}$ ($j=u,c,t$; $q=d,s$) are elements of the 
Cabibbo-Kobayashi-Maskawa matrix, and $c_i$ ($i=1,...,6$) the Wilson 
coefficients at the scale $\mu\simeq m_b$; $O_1^{u,c}$ and $O_2^{u,c}$ are 
current-current operators, $O_i$ ($i=3,..,6$) are penguin operators.

In the vacuum saturation approximation, the decay amplitudes can be written as
\begin{eqnarray}
<D^+D^-|H_W|B^0>&=&{G\over {\sqrt 2}}\left[V_{cd}V^*_{cb}a_1-V_{td}V^*_{tb}
\left(a_4+{{2a_6m^2_D}\over {(m_b-m_c)(m_c+m_d)}}\right)\right]\nonumber \\
&\times& <D^+|\bar c\gamma_\mu(1-\gamma_5)d|0><D^-|\bar b\gamma^\mu
(1-\gamma_5)b>|B^0>\ ,\end{eqnarray}
\begin{eqnarray}
<{D^*}^+{D^*}^-|H_W|B^0>&=&
{G\over {\sqrt 2}} (V_{cd}V^*_{cb}a_1-V_{td}V^*_{tb}a_4) \nonumber \\
&\times& <{D^*}^+|\bar c \gamma_\mu(1-\gamma_5)d|0> <{D^*}^-|\bar b 
\gamma^\mu(1-\gamma_5)c|B^0>\ ,
\end{eqnarray}
with parameters $a_1$,...,$a_6$ related to the Wilson coefficients $c_1$,...
,$c_6$ by the equations:
\begin{equation}
a_{2i-1}=c_{2i-1}+{{c_{2i}}\over{N_c}}\ ,\ a_{2i}=c_{2i}+{{c_{2i-1}}\over{N_c}}
\end{equation}
($i=1,2,3$ and $N_c$ is the number of colours).

In the NDR renormalization scheme at the next-to-leading order, using  
$N_c=3$ and the scale $\mu=m_b$, we get \cite{buras}
\begin{equation}
a_1=1.017\  ,\  a_2=0.175\  ,\  a_3=0.0013\  ,\  a_4=-0.030 , 
a_5=-0.0037 , a_6=-0.038 .
\end{equation}
Integrated and time-dependent asymmetries depend on the parameters of the 
Cabibbo-Kobayashi-Maskawa matrix and, unlike the $B^0\to J/\psi K_S$ mode, 
also on strong final state phases.
Various estimates, however, suggest that for $B$ transitions to charmed states 
such strong phases are small \cite{nardulli}, and therefore we neglect them 
in the analysis.
The asymmetry (\ref{a1t}) can be computed in a straightforward way; we omit 
here the lengthy expression, and only plot
in Fig.5 and Fig.6 the time dependent asymmetries for 
$B\to D^+D^-$ and $B\to {D^*}^+{D^*}^-$ respectively, for 
$\beta=22^\circ$ and the parameters of the CKM matrix $\rho=0.05$, 
$\eta=0.39$, $\gamma=83^\circ$; the corresponding error bars are also depicted
in the figures.

The analysis shows that a sensible bound on $s$ can be obtained from these 
decay channels only if tagging and reconstruction efficiencies at PEP II 
improve with respect to the estimates quoted in \cite{babar}.
For instance, an improvement by a factor of two will allow to put the bounds
$s\leq 0.18$ (from $B\to DD$) and $s\leq 0.22$ (from $B\to D^*D^*$)
in correspondence to $\Delta m\Delta t={\pi\over 2}$.

As for the hadronic channels $B^0\to \pi^+\pi^-$ and $B^0\to \rho^\pm 
\pi^\mp$, the $CP$ asymmetries depend on the strong phases due 
to the hadron rescattering in the final state. In any case, we found that 
the statistics expected for such decay modes does not allow to 
sensitively bound the $CPT$ violating parameter $s$.

\section{Testing direct $CPT$ violation effects}
So far we analyzed indirect $CPT$ violation effects. It is also possible to 
consider observables which are sensitive to direct violation effects. 
One of them is the asymmetry between the partial widths of charged $B$ mesons,
related to a $CPT$ violating term in the transition amplitudes
${<f|H_W|B^+> \neq<\bar f|H_W|B^->}$:
\begin{equation}
A={{\Gamma(B^+\to f)-\Gamma(B^-\to \bar f)}\over 
{\Gamma(B^+\to f)+\Gamma(B^-\to \bar f)}} \;\;\;.\label{dir}
\end{equation}
We do not parameterize this kind of violation, but just compute the 
expected error on the asymmetry (\ref{dir}). For the decay mode
$B^+\to\bar D^0 l^+ \nu_l$, $N=2.9\times 10^5$ identified events 
are expected per year \cite{babar}, so that the estimated 
error on (\ref{dir}) is
$\sigma =7.9\times 10^{-3}$; for $B^+\to\bar {D^0}^* l^+ \nu_l$, the number of 
expected events per year is $N=3.9\times 10^6$, and the estimated error is 
$\sigma =5.1\times10^{-3}$.

\section{Conclusions}

We investigated the possibility of testing $CPT$ symmetry in the neutral 
$B$ meson system, considering the experimental environment 
provided by the $B$-factory 
PEP II. We found that dileptons and lepton-hadron events, with the 
hadronic decay $B^0\to J/\psi K_S$, can be used to constrain
the $CPT$ violating parameter $s$. 

In particular, from equal and opposite sign dilepton production, one can derive
the bound 
$s\leq {\cal O}(10^{-2})$, which corresponds to a relative mass difference 
${|m_{B^0}-m_{\bar B^0}|\over m_{B_{average}}}\leq {\cal O}(10^{-16})$.

As for the hadronic decay $B^0\to J/\psi K_S$, the sensitivity of PEP II is 
up to $s\simeq {\cal O}(10^{-1})$.
Our analysis, though being quite different, confirms the estimate 
in \cite{kostelecky} about the order of $CPT$ violating effects that can be 
constrained at PEP II.

Other channels, such as $B^0\to D^+D^-$ and $B^0\to {D^*}^+{D^*}^-$, might 
allow a sensible bounding of $s$ if tagging and reconstruction efficiencies 
improve at least by a factor of two with respect to the current expectations. 
Charmless decays of the type $B^0\to \pi^+\pi^-$, 
$B^0\to \rho^\pm\pi^\mp$, turn out to be 
unsuitable for testing $CPT$.

\vskip 1cm
\noindent {\bf Acknowledgments}\\
\noindent {We are indebted with G.Nardulli for interesting 
discussions and suggestions. We also acknowledge F. De Fazio for discussions.}

\newpage
\appendix
\section{Case $C=+1$}

For the sake of completeness, we report the relevant formulae corresponding to 
a $C$ even $B^0$-$\bar B^0$ state:
\begin{eqnarray}
P_+^{-+}&\propto& {1\over 4}(|A_l|^2)^2 {{e^{-\Gamma (t_1+t_2)}}\over
{4(1+s^2)^2}}
\{1+\cos \Delta m(t_1+t_2) \nonumber \\
&+&s^2[1+2\cos \Delta mt_1+2\cos \Delta mt_2-
\cos\Delta m(t_1-t_2)]\}\ .
\end{eqnarray}
\begin{eqnarray}
P^{\pm \pm}_+(t_1, t_2)&\propto& (|A_l|^2)^2{{e^{-\Gamma (t_1+t_2)}}\over
{4(1+s^2)^2}}\{1-\cos \Delta m(t_1+t_2)\nonumber \\
&+&s^2[3-2\cos \Delta mt_1-2\cos
\Delta mt_2+\cos \Delta m(t_1+t_2)]\}\ .
\end{eqnarray}
\begin{equation}
n_+^{-+}={1\over {2(1+s^2)^2}} \left[{{x_d^4+x_d^2+2}\over {(1+x_d^2)
^2}}+ s^2 {{4+x_d^2}\over {1+x_d^2}}\right]
\end{equation}
\begin{equation}
n_+^{++}=n_+^{--}={1\over {4(1+s^2)}}
\left[ {{x_d^4+3x_d^2}\over {(1+x_d)^2}}
+s^2 {{3x_d^2}\over {1+x_d^2}}\right]
\end{equation}
\begin{eqnarray}
n^{-+}_+(\Delta t)&=&{{e^{-\Gamma |\Delta t|}}\over {2(1+s^2)^2}}
\bigg\{ {{2+x_d^2}\over {1+x_d^2}}+s^2 \bigg[1-\cos\Delta m\Delta t+
{2\over {\sqrt{1+x_d^2/4}}}(\sin(\Delta m|\Delta t|-\alpha)
\nonumber \\
&+&\cos(\Delta m|\Delta t|-\alpha)\bigg]\bigg\}
\end{eqnarray}
\begin{eqnarray}
n^{++}_+(\Delta t)&=&{{e^{-\Gamma |\Delta t|}}\over {4(1+s^2)^2}}
\bigg\{ {{x_d^2}\over {1+x_d^2}}+s^2 \bigg[3+\cos\Delta m\Delta t+
{2\over {\sqrt{1+x_d^2/4}}}(\sin(\Delta m|\Delta t|-\alpha)
\nonumber \\
&+&\cos(\Delta m|\Delta t|-\alpha))\bigg]\bigg\}\ ,
\end{eqnarray}
with $\alpha=\arctan{{x_d}\over 2}$.

\clearpage

\hskip 5 cm {\bf FIGURE CAPTIONS}
\vskip 1 cm
 
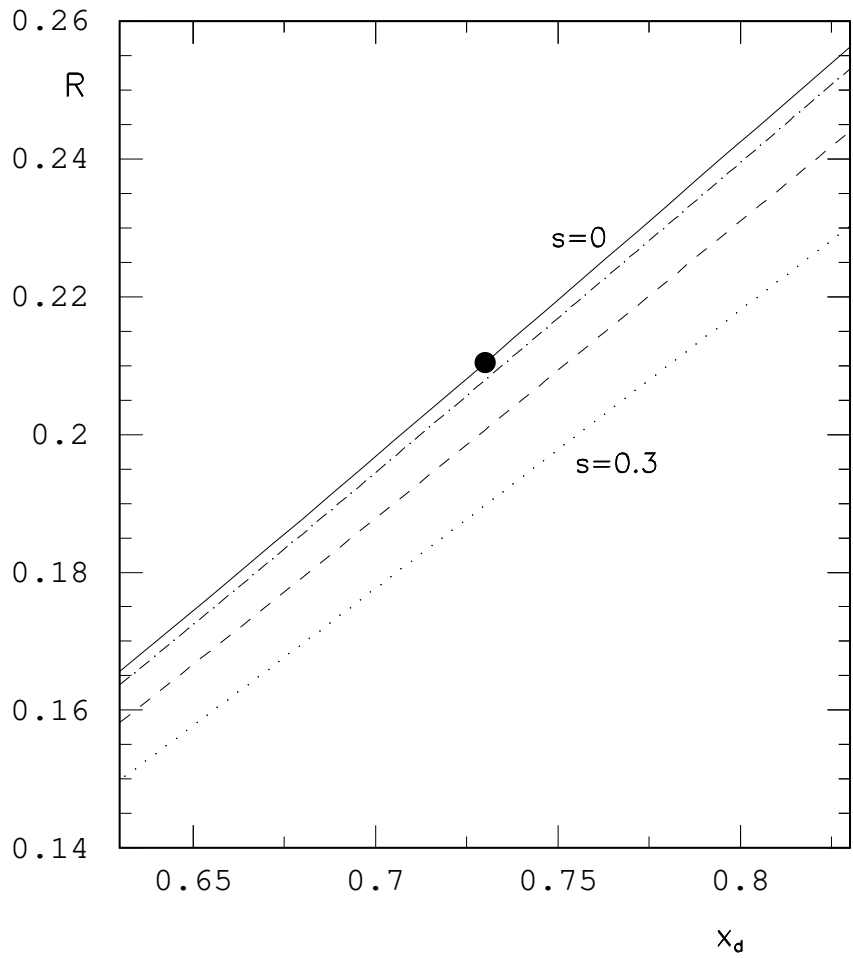
\noindent {\bf Fig. 1}\\
\noindent
The ratio $R$ in (\ref{R}) versus $x_d$ and $s$. 
The continuous line corresponds to $s=0$, the dashed-dotted line to $s=0.1$, 
the dashed line to $s=0.2$, the dotted line to $s=0.3$.
\vspace{4mm}

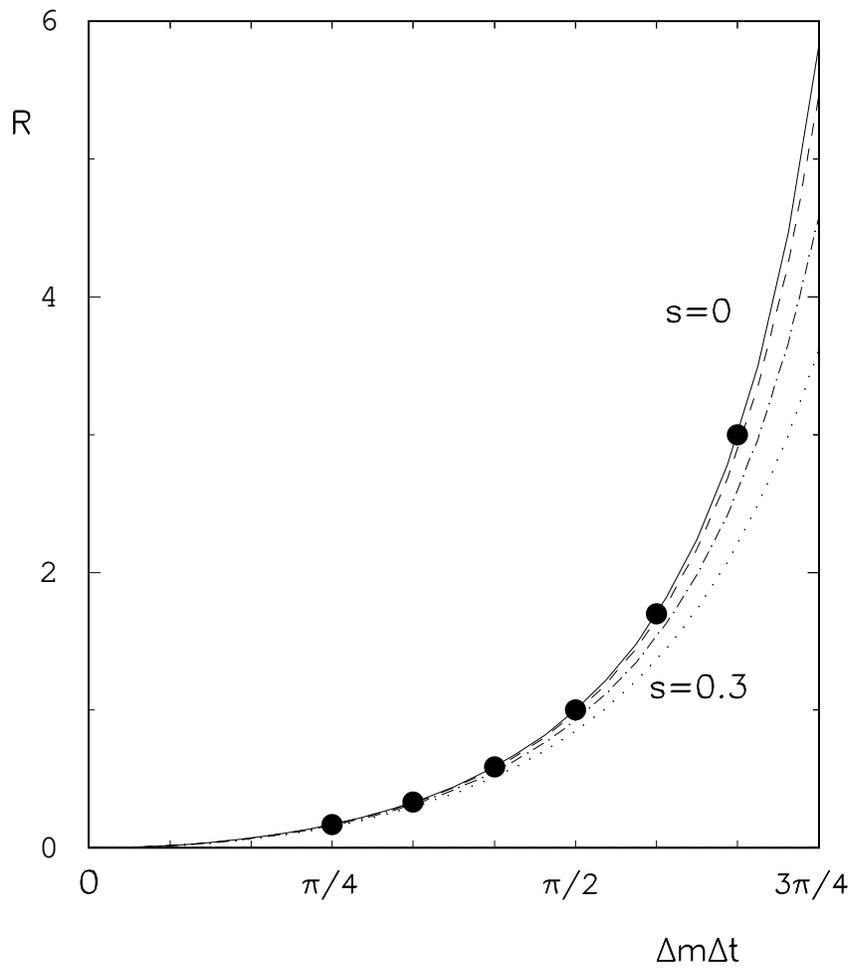
\noindent {\bf Fig. 2}\\
\noindent
Time dependent results for $R$ (eq.(\ref{Rdt})).
Notations as in fig.1. The expected statistical error is smaller than the size 
of the dots for the different values of $\Delta m\Delta t$.
\vspace{4mm}

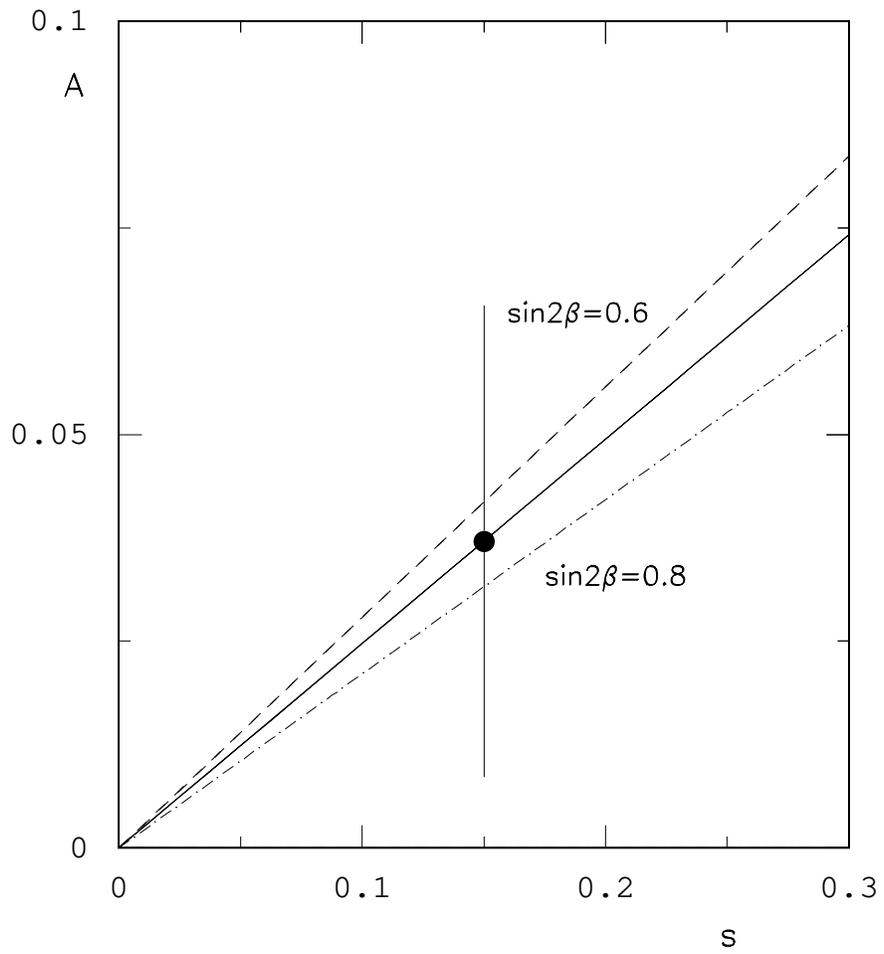
\noindent {\bf Fig. 3}\\
\noindent
Integrated asymmetry (\ref{abk}) for the channel $B^0\to J/\psi K_S$. The 
three lines correspond to the values of the $\beta$ angle: 
$\beta= 18^\circ$ (dashed-dotted line), $\beta= 22^\circ$
(solid line) and $\beta=26^\circ$ (dashed line). Statistical 
errors show the minimum value of $s$ accessible from this mode.
\vspace{4mm}

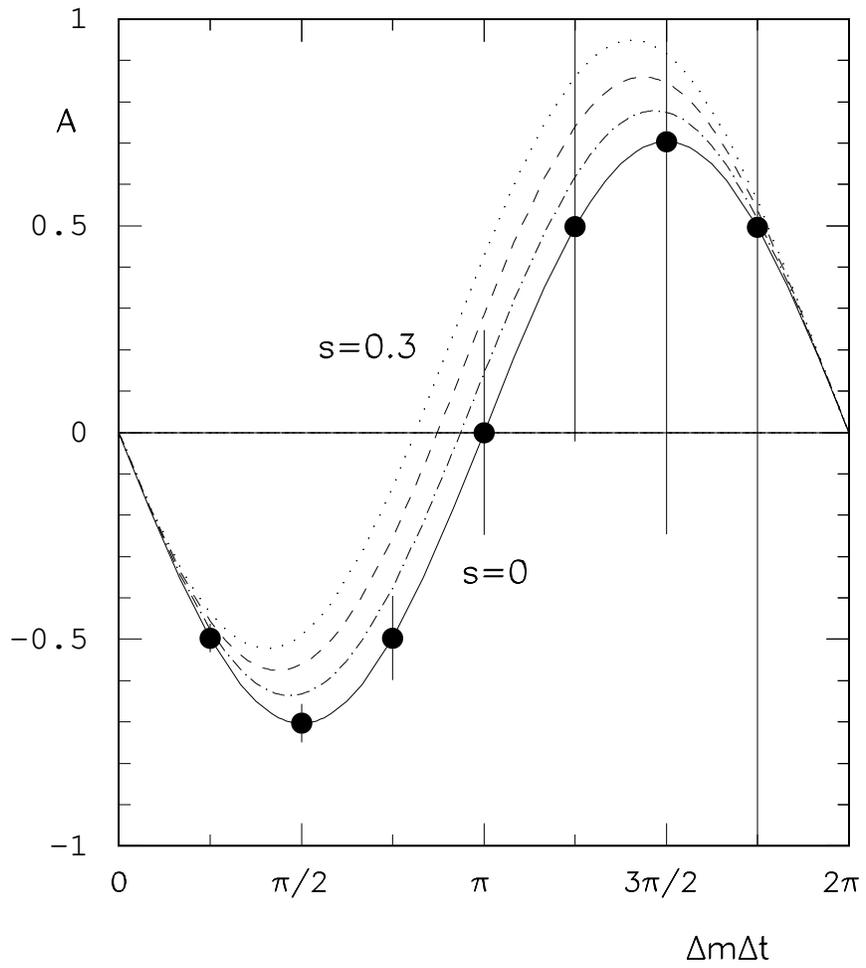
\noindent {\bf Fig. 4}\\
\noindent
Time dependent asymmetry (\ref{abkt}) for the channel $B^0\to J/\psi K_S$.
The lines correspond to the values $s=0$ (solid line), $s=0.1$ 
(dashed-dotted line), $s=0.2$ (dashed line) and $s=0.3$ (dotted line).
The expected statistical error for different values of $\Delta m\Delta t$ 
is also shown.
\vspace{4mm}

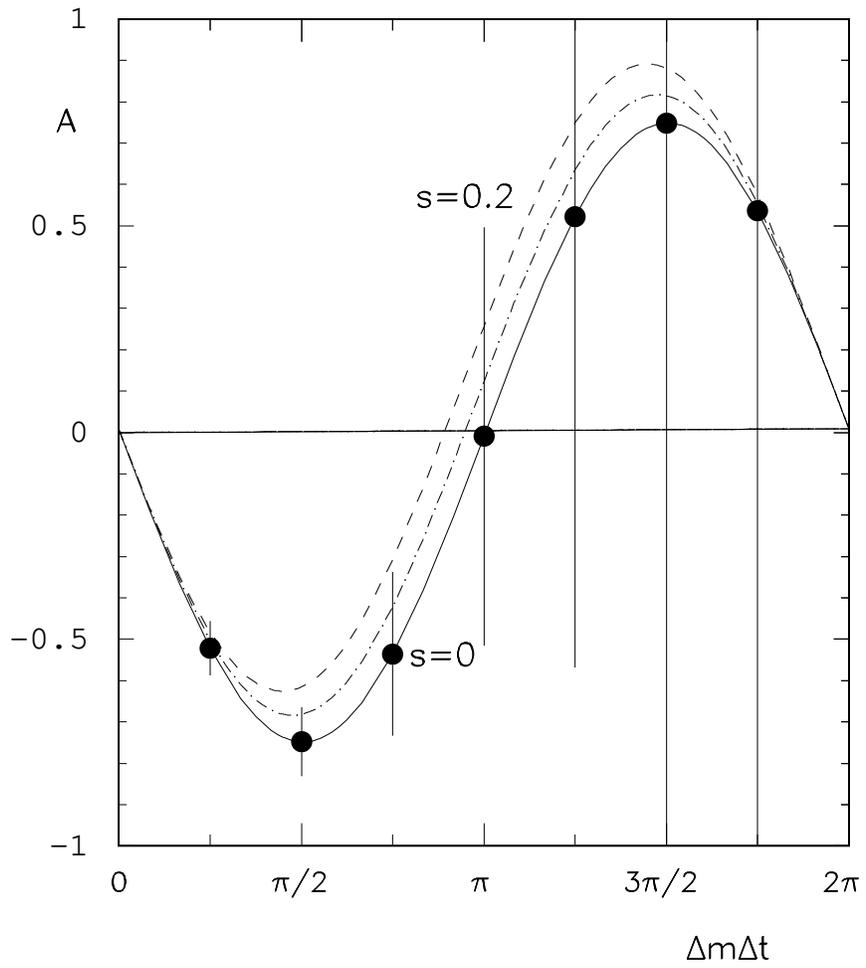
\noindent {\bf Fig. 5}\\
\noindent
Time dependent asymmetry (\ref{a1t}) for the channel $B^0\to D^+D^-$.
The lines correspond to the values $s=0$ (solid line), $s=0.1$ (dashed-dotted
line) and $s=0.2$ (dashed line).
\vspace{4mm}

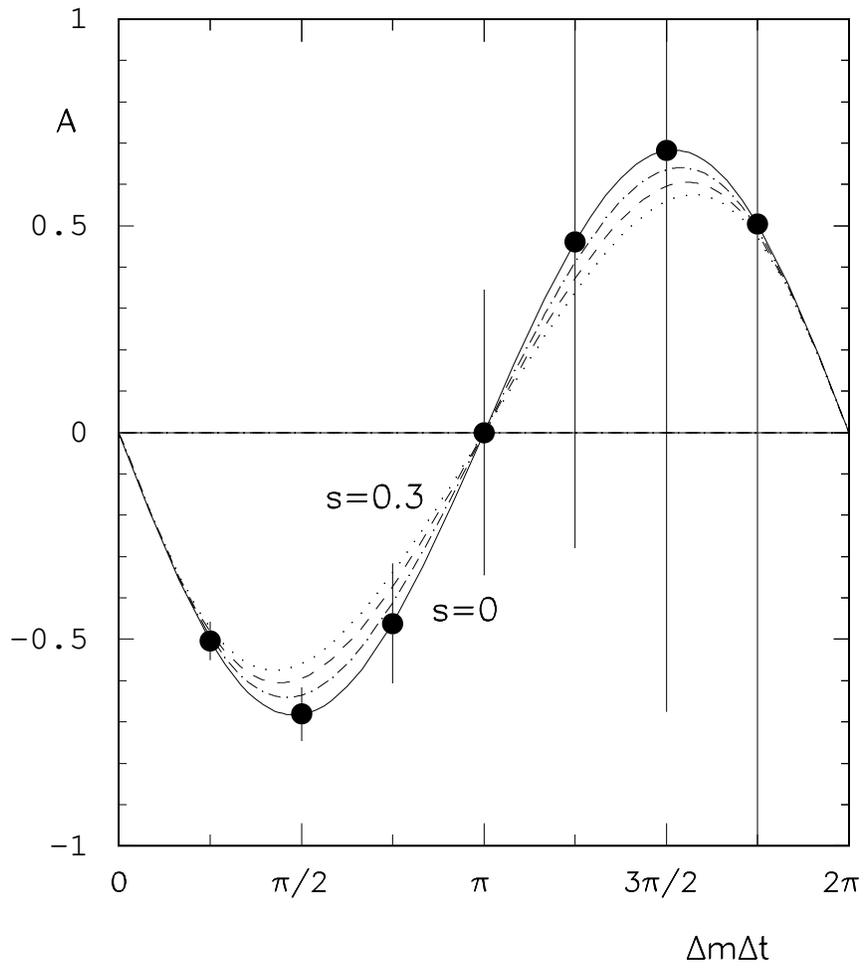
\noindent {\bf Fig. 6}\\
\noindent
Time dependent asymmetry (\ref{a1t}) for the channel $B^0\to {D^*}^+{D^*}^-$.
The lines correspond to the values $s=0$ (solid line), $s=0.1$ (dashed-dotted 
line), $s=0.2$ (dashed line) and $s=0.3$ (dotted line).

\end{document}